\documentclass[eqsecnum,showpacs,preprint,amsmath,aps,nofootinbib]{revtex4}  %

\begin{document}
\title{One-loop analysis with nonlocal boundary conditions}

\author{Giampiero Esposito ORCID: 0000-0001-5930-8366}
\email[E-mail: ]{gesposit@na.infn.it}
\affiliation{Istituto Nazionale di Fisica Nucleare, Sezione di
Napoli, Complesso Universitario di Monte S. Angelo, 
Via Cintia Edificio 6, 80126 Napoli, Italy}

\author{Elisabetta Di Grezia ORCID: 0000-0003-0422-4573}
\email[E-mail: ]{digrezia@na.infn.it}
\affiliation{Istituto Nazionale di Fisica Nucleare, Sezione di
Napoli, Complesso Universitario di Monte S. Angelo, 
Via Cintia Edificio 6, 80126 Napoli, Italy}

\date{\today}

\begin{abstract}
In the eighties, Schr\"{o}der studied a quantum mechanical model where the stationary states of
Schr\"{o}dinger's equation obey nonlocal boundary conditions on a circle in the plane. For such a problem, 
we perform a detailed one-loop calculation for three choices of the kernel characterizing the nonlocal 
boundary conditions. In such cases, the $\zeta(0)$ value is found to coincide with the one
resulting from Robin boundary conditions. The detailed technique here developed may be useful for
studying one-loop properties of quantum field theory and quantum gravity if nonlocal boundary conditions
are imposed.
\end{abstract}


\maketitle

\section{Introduction}

In the late eighties, motivated by physical models of Bose condensation and mathematical study of
Schr\"{o}dinger operators, the work in Ref. \cite{Schroder} studied spectral properties of the Laplace operator
with nonlocal boundary conditions. Within this framework, on considering the region
\begin{equation}
B_{R} \equiv \left \{ x,y: x^{2}+y^{2} \leq R^{2} \right \} ,
\label{(1.1)}
\end{equation}
one builds, out of a function $q$ which is both Lebesgue summable and square-integrable on the 
real line, the periodic function
\begin{equation}
q_{R}(x) \equiv {1 \over 2\pi R} \sum_{l=-\infty}^{\infty}e^{i lx/R} 
\int_{-\infty}^{\infty}e^{-ily/R}q(y)dy,
\label{(1.2)}
\end{equation}
which has period $2 \pi R$ and approaches $q$ if $R \rightarrow \infty$. On using polar coordinates 
$(r,\varphi)$, the nonlocal boundary-value problem studied in Ref. \cite{Schroder} reads as
\begin{equation}
-\left ({\partial^{2}\over \partial r^{2}}+{1 \over r}{\partial \over \partial r}
+{1 \over r^{2}}{\partial^{2}\over \partial \varphi^{2}}\right)u=Eu, \; \forall r<R,
\label{(1.3)}
\end{equation}
\begin{equation}
{\partial u \over \partial r}+R \int_{-\pi}^{\pi}
q_{R}(R(\varphi-\theta)) \; u(R,\theta)d \theta=0.
\label{(1.4)}
\end{equation}
The resulting spectrum has both a positive ($E>0$) and a negative ($E<0$) part. For $E>0$, the 
solutions regular at the origin $r=0$ are factorized in the form
\begin{equation}
u_{l,E}(r,\varphi)=J_{l}(r \sqrt{E}) e^{il \varphi}, \;
l \in Z,
\label{(1.5)}
\end{equation}
the $J_{l}$ being Bessel functions of first kind. On defining the dimensionless variable
$h \equiv R \sqrt{E}$, Eq. (1.4) takes eventually the form \cite{Schroder}
\begin{equation}
G_{l}(h) \equiv h J_{l}'(h)+R {\tilde q} \left({l \over R}\right) J_{l}(h)=0,
\label{(1.6)}
\end{equation}
where ${\tilde q}$ is the Fourier transform of $q$, i.e. [1]
\begin{equation}
{\tilde q} \left({l \over R}\right) = \int_{-\infty}^{\infty}
q(x) e^{-i{l \over R}x} \; dx.
\label{(1.7)}
\end{equation}
We note from (1.6) that ${\tilde q}$ must have dimension ${\rm length}^{-1}$, and 
hence $q$ must have dimension ${\rm length}^{-2}$.

In the present paper, we have tried to work out the one-loop properties pertaining to the
problem defined by Eqs. (1.3) and (1.4). In the physics-oriented literature, one-loop
calculations are more frequently performed in the case of quantum field theories,
but the quantum mechanical framework is already of interest \cite{Barvinsky}, and may provide
valuable information on the behaviour of solutions of elliptic equations under
a scale dilation. Such a property is neatly described by the $\zeta(0)$ value, where
$\zeta$ is the spectral (or generalized) $\zeta$-function of the elliptic operator $A$
under consideration, defined by
\begin{equation}
\zeta_{A}(s) \equiv {\rm Tr}_{L^{2}}(A^{-s})= \sum_{n}\lambda_{n}^{-s},
\label{(1.8)}
\end{equation}
where each eigenvalue $\lambda_{n}$ is counted with its degeneracy.
For this purpose, section $2$ outlines the analytic technique of Ref. \cite{Moss} and its
application to our boundary-value problem (1.3) and (1.4). Section $3$ evaluates the first $3$ 
sums that contribute to the $\zeta(0)$ value, whereas section $4$ studies the fourth and last
sum contributing to $\zeta(0)$, for three choices of the function $q$ and hence of
the dimensionless coefficients $\beta_{l}$ from Eqs. (1.6) and (1.7), i.e.
\begin{equation}
\beta_{l} \equiv R {\tilde q} \left({l \over R} \right).
\label{(1.9)}
\end{equation}
Concluding remarks are presented in section $5$, whereas relevant details are given
in the appendix.

\section{$\zeta(0)$ value from the $\zeta$-function at large $h$}

For the Laplacian $A$ on the left-hand side of Eq. (1.3), the associated heat equation has a
heat kernel $K(\xi,\eta;t)$, whose diagonal $K(\xi,\xi;t)$ yields, upon integration over
the whole region $B_{R}$ in (1.1), the integrated heat kernel (for gauge theories, the
trace to be integrated is instead the fibre trace of the heat-kernel diagonal) 
\begin{equation}
K(t)=\int_{B_{R}}K(\xi,\xi;t)={\rm Tr}_{L^{2}}e^{-tA},
\label{(2.1)}
\end{equation}
which has, as $t \rightarrow 0^{+}$, the asymptotic expansion \cite{Moss,Esposito98,Gilkey}
\begin{equation}
K(t) \sim \sum_{n=0}^{\infty}B_{n}t^{{n \over 2}-1}.
\label{(2.2)}
\end{equation}
In our $2$-dimensional region $B_{R}$, the method used in Ref. \cite{Moss} considers the so-called
spectral $\zeta$-function at large $h$, i.e. ($\lambda_{n}$ and $h$ 
being dimensionless in (2.3))
\begin{equation}
\zeta(s,h^{2}) \equiv \sum_{n}(\lambda_{n}+h^{2})^{-s},
\label{(2.3)}
\end{equation}
which is related to the integrated heat kernel (2.1) by the identity
\begin{equation}
\int_{0}^{\infty}t e^{-h^{2}t}K(t)dt=\Gamma(2)\zeta(2,h^{2}).
\label{(2.4)}
\end{equation}
If one now inserts into the left-hand side of (2.4) the asymptotic expansion (2.2), one finds
\begin{equation}
\Gamma(2)\zeta(2,h^{2}) \sim \sum_{n=0}^{\infty}B_{n}
\Gamma \left(1+{n \over 2}\right)h^{-n-2}.
\label{(2.5)}
\end{equation}
On the other hand, on considering the equation (1.6), which is the equation obeyed by the
eigenvalues $E={h^{2}\over R^{2}}$ by virtue of the boundary conditions, one has also the identity
(see \cite{Moss} and our appendix)
\begin{equation}
\Gamma(2) \zeta(2,h^{2})=\sum_{l}^{\infty}(-N_{l})
\left({1 \over 2h}{d \over dh}\right)^{2}\log \Bigr[(ih)^{-l}G_{l}(ih)\Bigr],
\label{(2.6)}
\end{equation}
where the degeneracy $N_{l}$ equals $2$, as is clear from (1.5) (for each value of $l$,
there exist $2$ linearly independent eigenfunctions proportional to $\cos(l \varphi)$ and
$\sin (l \varphi)$, respectively). In light of (2.5) and (2.6), the desired 
$\zeta(0)$ is the coefficient of $h^{-4}$ in the asymptotic expansion of the right-hand
side of Eq. (2.6), because $\zeta(0)=B_{2}$ from the formulae [4]
\begin{eqnarray}
\zeta(s) & = & {1 \over \Gamma(s)} \int_{0}^{\infty} t^{s-1}K(t) \; dt 
\nonumber \\
& \sim & {1 \over \Gamma(s)}
\left[\sum_{n=0}^{\infty}B_{n} \int_{0}^{1}t^{{n \over 2}+s-2}dt
+\int_{1}^{\infty}t^{s-1}K(t) \; dt \right].
\label{(2.7)}
\end{eqnarray}

In the course of performing sums over all positive and negative values of $l$, it is helpful
to exploit the identity
\begin{equation}
J_{-l}(z)=(-1)^{l}J_{l}(z),
\label{(2.8)}
\end{equation}
as well as the even nature of $\beta_{l}$ as a function of $l$ (see section $4$). This implies that
a real root of $G_{l}$ with positive $l$ is also a real root of $G_{l}$ with negative $l$,
because
\begin{equation}
G_{-l}(z)=(-1)^{l}G_{l}(z).
\label{(2.9)}
\end{equation}
We can therefore limit ourselves to summing over positive values of $l$, writing that, in (2.6),
\begin{equation}
\sum_{l}^{\infty} = 2 \sum_{l=1}^{\infty} 
+ \; {\rm contribution} \; {\rm of} \; (l=0) 
= 2 \sum_{l=0}^{\infty}- \; {\rm contribution} \; {\rm of} \; (l=0).
\label{(2.10)}
\end{equation}

Following Ref. \cite{Moss}, we use in (2.6) and (2.10) the uniform asymptotic expansion of $J_{l}$ and 
its first derivative $J_{l}'$, which involve the polynomials $u_{k}$ and $v_{k}$ 
occurring below and in the appendix. On denoting by $C$ a constant, and defining
$\alpha_{l}(ih) \equiv \sqrt{l^{2}+h^{2}}$, we obtain
\begin{eqnarray}
\; & \; & \log \Bigr[(ih)^{-l}G_{l}(ih)\Bigr] \sim
C -l \log(l+\alpha_{l})+{1 \over 2}\log(\alpha_{l})+\alpha_{l} 
\nonumber \\
& + & \log \left[1+{(b_{1}+\beta_{l})\over \alpha_{l}}
+{(b_{2}+\beta_{l}a_{1})\over \alpha_{l}^{2}}
+{(b_{3}+\beta_{l}a_{2})\over \alpha_{l}^{3}}
+{\rm O}(\alpha_{l}^{-4}) \right],
\label{(2.11)}
\end{eqnarray}
where, having defined the variable
\begin{equation}
\tau \equiv {l \over \alpha_{l}} = {l \over \sqrt{l^{2}+h^{2}}},
\label{(2.12)}
\end{equation}
and exploiting the following among the many Debye-Olver polynomials \cite{Olver}:
\begin{equation}
u_{0}(\tau)=1, \; u_{1}(\tau)={\tau \over 8}-{5 \over 24}\tau^{3},
\label{(2.13)}
\end{equation}
\begin{equation}
u_{2}(\tau)={9 \over 128}\tau^{2}-{77 \over 192}\tau^{4}
+{385 \over 1152}\tau^{6},
\label{(2.14)}
\end{equation}
\begin{equation}
v_{0}(\tau)=1, \; v_{1}(\tau)=-{3 \over 8}\tau+{7 \over 24}\tau^{3},
\label{(2.15)}
\end{equation}
\begin{equation}
v_{2}(\tau)=-{15 \over 128}\tau^{2}+{99 \over 192}\tau^{4}-{455 \over 1152}\tau^{6},
\label{(2.16)}
\end{equation}
\begin{equation}
v_{3}(\tau)=-{105 \over 1024}\tau^{3}+{5577 \over 5120}\tau^{5}-{6545 \over 3072}\tau^{7}
+{95095 \over 82944}\tau^{9},
\label{(2.17)}
\end{equation}
the polynomials $a_{1},a_{2},b_{1},b_{2},b_{3}$ are evaluated according to the definition
\begin{equation}
a_{k}(\tau)={u_{k}(\tau)\over \tau^{k}}, \; b_{k}(\tau)={v_{k}(\tau)\over \tau^{k}},
\label{(2.18)}
\end{equation}
which implies the simple but very helpful relations
\begin{equation}
{u_{k}\over l^{k}}={a_{k}\over (\alpha_{l})^{k}}, \;
{v_{k}\over l^{k}}={b_{k}\over (\alpha_{l})^{k}}.
\label{(2.19)}
\end{equation}

\section{Contributions independent of $\beta_{l}$}

By virtue of (2.6) and (2.11) the three contributions independent of $\beta_{l}$ are
obtained by applying twice the operator ${1 \over 2h}{d \over dh}$ to the first line
on the right-hand side of (2.11). For this purpose, we need the following identities:
\begin{equation}
{1 \over 2h}{d \alpha_{l}\over dh}={1 \over 2 \alpha_{l}},
\label{(3.1)}
\end{equation}
\begin{equation}
\left({1 \over 2h}{d \over dh}\right)^{2}\log(l+\alpha_{l})=-{1 \over 4}(l+\alpha_{l})^{-2}
\Bigr[2 \alpha_{l}^{-2}+l \alpha_{l}^{-3}\Bigr],
\label{(3.2)}
\end{equation}
\begin{equation}
\left({1 \over 2h}{d \over dh}\right)^{2}\log(\alpha_{l})=-{1 \over 2}\alpha_{l}^{-4},
\label{(3.3)}
\end{equation}
\begin{equation}
\left({1 \over 2h}{d \over dh}\right)^{2}\alpha_{l}=-{1 \over 4}\alpha_{l}^{-3}.
\label{(3.4)}
\end{equation}
Thus, upon applying the split (2.10), the terms independent of $\beta_{l}$ are obtained by 
taking twice (from the factor $2$ multiplying $\sum_{l=0}^{\infty}$ in (2.10)) 
the following sums:
\begin{equation}
\sigma_{1} = 2 \sum_{l=0}^{\infty}
\left({1 \over 2h}{d \over dh}\right)^{2}
l \log(l+\alpha_{l}) 
=-{1 \over 2}\sum_{l=0}^{\infty}l (l+\alpha_{l})^{-2}
\Bigr[2 \alpha_{l}^{-2}+l \alpha_{l}^{-3}\Bigr],
\label{(3.5)}
\end{equation}
\begin{equation}
\sigma_{2}=-\sum_{l=0}^{\infty}
\left({1 \over 2h}{d \over dh}\right)^{2}\log(\alpha_{l})
={1 \over 2}\sum_{l=0}^{\infty}\alpha_{l}^{-4},
\label{(3.6)}
\end{equation}
\begin{equation}
\sigma_{3}=-2 \sum_{l=0}^{\infty}
\left({1 \over 2h}{d \over dh}\right)^{2}\alpha_{l}
={1 \over 2}\sum_{l=0}^{\infty}\alpha_{l}^{-3}.
\label{(3.7)}
\end{equation}

\subsection{Contribution of $\sigma_{1}$}

The sums (3.5)-(3.7) can be studied in a thorough way with the help of the Euler-Maclaurin summation 
formula \cite{Wong}. This states that, if $f$ is a real- or complex-valued function defined on
$[0,\infty)$, and if its derivatives of even order are absolutely integrable on $(0,\infty)$,
one has, for $n=1,2,...$,
\begin{equation}
\sum_{l=0}^{n}f(l)-\int_{0}^{n}f(y)dy={1 \over 2}[f(0)+f(n)]
+\sum_{s=1}^{m-1}{{\tilde B}_{2s}\over (2s)!}
\Bigr[f^{(2s-1)}(n)-f^{(2s-1)}(0)\Bigr]+R_{m}(n),
\label{(3.8)}
\end{equation}
where the Bernoulli numbers ${\tilde B}_{s}$ are defined by the expansion
\begin{equation}
{t \over (e^{t}-1)}=\sum_{s=0}^{\infty}{\tilde B}_{s}{t^{s}\over s!}, \;
|t| < 2 \pi,
\label{(3.9)}
\end{equation}
while the remainder $R_{m}(n)$ is majorized according to \cite{Wong}
\begin{equation}
|R_{m}(n)| \leq \Bigr(2-2^{1-m}\Bigr) {|{\tilde B}_{2m}|\over (2m)!}
\int_{0}^{n} |f^{(2m)}(y)|dy.
\label{(3.10)}
\end{equation}
As $n$ approaches $\infty$, Eq. (3.8) provides a very useful asymptotic expansion for the
desired sum of the series, i.e.
\begin{equation}
\sum_{l=0}^{\infty}f(l) \sim \int_{0}^{\infty}f(y)dy+{1 \over 2}f(0)
+\sum_{s}{{\tilde B}_{2s}\over (2s)!}
\Bigr[f^{(2s-1)}(\infty)-f^{(2s-1)}(0)\Bigr].
\label{(3.11)}
\end{equation}
The integral on the right-hand side of (3.11) can be evaluated or studied in a qualitative way,
while the derivatives of odd order at $0$ and at $\infty$ can be obtained in a systematic way.
We refer the reader to the last chapter of the book by Hardy \cite{Hardy} for a thorough analysis
of the Euler-Maclaurin formula.

For our purposes, after having re-expressed $\sigma_{1}$ in the form
\begin{equation}
\sigma_{1}=\sum_{l=0}^{\infty}[F(l;h)+H(l;h)],
\label{(3.12)}
\end{equation}
where we have set
\begin{equation}
F(l;h) \equiv -l(l^{2}+h^{2})^{-1}(l+\sqrt{l^{2}+h^{2}})^{-2},
\label{(3.13)}
\end{equation}
\begin{equation}
H(l;h) \equiv -{l^{2}\over 2}(l^{2}+h^{2})^{-{3 \over 2}} 
(l+\sqrt{l^{2}+h^{2}})^{-2},
\label{(3.14)}
\end{equation}
we now take the limit as $n \rightarrow \infty$ in Eq. (3.8), with $f(l)$ replaced by
$F(l;h)+H(l;h)$. We then find that, in (3.11), only the first derivative of $F$ at $l=0$ gives 
a contribution proportional to $h^{-4}$, and indeed equal to
\begin{equation}
\delta_{1}={1 \over 2!}{\tilde B}_{2}(-F'(0;h))={1 \over 12}h^{-4}.
\label{(3.15)}
\end{equation}

\subsection{Contributions of $\sigma_{2}$ and $\sigma_{3}$}

We now rely again upon the limit as $n \rightarrow \infty$ of Eq. (3.8).
By virtue of (3.11), only half the value at $l=0$ of ${1 \over 2}\alpha_{l}^{-4}$ 
contributes to the $h^{-4}$ term in $\sigma_{2}$, i.e.
\begin{equation}
\delta_{2}={1 \over 2}{1 \over 2}h^{-4}={1 \over 4}h^{-4},
\label{(3.16)}
\end{equation}
whereas $\sigma_{3}$ gives a vanishing contribution to the term 
proprtional to $h^{-4}$
\begin{equation}
\delta_{3}=0.
\label{(3.17)}
\end{equation}

\section{Contribution from the $\beta_{l}$ coefficients}

We now aim at studying the contribution of the second line of the asymptotic expansion 
(2.11) to Eq. (2.6). For this purpose, on the one hand we denote by $\Omega$ all terms added 
to $1$ within the square brackets in (2.11), and consider the asymptotic expansion
\begin{equation} 
\log(1+\Omega) \sim \Omega-{\Omega^{2}\over 2}+{\Omega^{3} \over 3}
+{\rm O}(\Omega^{4}) 
\sim {\kappa_{1} \over \alpha_{l}}
+{\kappa_{2}\over (\alpha_{l})^{2}}
+{\kappa_{3}\over (\alpha_{l})^{3}}+{\rm O}(\alpha_{l}^{-4}),
\label{(4.1)}
\end{equation}
where
\begin{equation}
\kappa_{1} \equiv b_{1}+\beta_{l},
\label{(4.2)}
\end{equation}
\begin{equation}
\kappa_{2} \equiv (b_{2}+\beta_{l}a_{1})-{1 \over 2}(b_{1}+\beta_{l})^{2},
\label{(4.3)}
\end{equation}
\begin{equation}
\kappa_{3} \equiv b_{3}+\beta_{l}a_{2}-(b_{1}+\beta_{l})(b_{2}+\beta_{l}a_{1})
+{1 \over 3}(b_{1}+\beta_{l})^{3}.
\label{(4.4)}
\end{equation}

On the other hand, it is clear that no further progress can be made without explicit 
forms of the $\beta_{l}$ coefficients. For example, we find from (1.7) and (1.9)
\begin{equation}
q(x)={1 \over R^{2}}e^{-{x^{2}\over R^{2}}} \Longrightarrow 
\beta_{l}={1 \over \sqrt{2}}e^{-{l^{2}\over 4}},
\label{(4.5)}
\end{equation}
\begin{equation}
q(x)=\sqrt{\pi \over 2}{1 \over R^{2}}e^{-{|x|\over R}} \Longrightarrow
\beta_{l}={1 \over (1+l^{2})},
\label{(4.6)}
\end{equation}
\begin{equation}
q(x)=\sqrt{2 \over \pi}{1 \over x^{2}} \Longrightarrow 
\beta_{l}=-l {\rm sgn}(l),
\label{(4.7)}
\end{equation}
\begin{equation}
q(x)=\sqrt{2 \over \pi}{1 \over (R^{2}+x^{2})}
\Longrightarrow \beta_{l}=e^{-|l|}.
\label{(4.8)}
\end{equation}

Here we consider first the choice of $\beta_{l}$ in Eq. (4.6), and exploit the
formulae (2.6), (2.11)-(3.1) and (4.1)-(4.4), arriving therefore at the sums
(see details below)
\begin{eqnarray}
({\hat \sigma}_{4})_{I} & \equiv & -2 \sum_{l=0}^{\infty}
\left({1 \over 2h}{d \over dh}\right)^{2}{\kappa_{1}\over \alpha_{l}} 
\nonumber \\
& = & {9 \over 16} \sum_{l=0}^{\infty} \alpha_{l}^{-5}
-{35 \over 16} \sum_{l=0}^{\infty} l^{2}\alpha_{l}^{-7}
-{3 \over 2} \sum_{l=0}^{\infty}{\alpha_{l}^{-5}\over (l^{2}+1)},
\label{(4.9)}
\end{eqnarray}
\begin{equation}
({\hat \sigma}_{4})_{II}  \equiv  -2 \sum_{l=0}^{\infty}
\left({1 \over 2h}{d \over dh}\right)^{2}
{\kappa_{2}\over (\alpha_{l})^{2}}, 
\label{(4.10)}
\end{equation}
supplemented, in principle, by infinitely many other terms, i.e.
\begin{equation}
({\hat \sigma}_{4})_{m} \equiv
-2 \sum_{l=0}^{\infty}
\left({1 \over 2h}{d \over dh}\right)^{2}
{\kappa_{m}\over (\alpha_{l})^{m}}, \;
\forall m=3,4,...,\infty .
\label{(4.11)}
\end{equation}
In the formula (4.10), it is helpful to use (4.3) where we re-express $a_{1},b_{1}$ 
and $b_{2}$ in the form
\begin{equation}
a_{1}=\sum_{r=0}^{1}a_{1r}\left({l \over \alpha_{l}}\right)^{2r}, \;
b_{1}=\sum_{r=0}^{1}b_{1r}\left({l \over \alpha_{l}}\right)^{2r}, \;
b_{2}=\sum_{r=0}^{2}b_{2r}\left({l \over \alpha_{l}}\right)^{2r},
\label{(4.12)}
\end{equation}
where the numerical coefficients $a_{1r},b_{1r}$ and $b_{2r}$ can be read off from 
(2.13), (2.15) and (2.16). Thus, a patient calculation shows that
(the superscript $(l)$ denotes here dependence on $l$) 
\begin{equation}
{\kappa_{2}\over (\alpha_{l})^{2}}=\sum_{r=0}^{2}
\kappa_{2r}^{(l)} l^{2r}\alpha_{l}^{-2r-2},
\label{(4.13)}
\end{equation}
where
\begin{equation}
\kappa_{20}^{(l)}=b_{20}+\beta_{l}(a_{10}-b_{10})-{1 \over 2}\left((b_{10})^{2}+(\beta_{l})^{2}\right)
=-{3 \over 16}+{1 \over 2}\beta_{l}(1-\beta_{l}),
\label{(4.14)}
\end{equation}
\begin{equation}
\kappa_{21}^{(l)}=b_{21}+\beta_{l}(a_{11}-b_{11})-b_{10}b_{11}={5 \over 8}
-{1 \over 2}\beta_{l},
\label{(4.15)}
\end{equation}
\begin{equation}
\kappa_{22}^{(l)}=b_{22}-{1 \over 2}(b_{11})^{2}=-{7 \over 16},
\label{(4.16)}
\end{equation}
and hence, by repeated application of (3.1), we obtain from (4.10) and (4.13)
\begin{eqnarray}
({\hat \sigma}_{4})_{II}&=& -2 \sum_{l=0}^{\infty} \sum_{r=0}^{2} (r+1)(r+2)
\kappa_{2r}^{(l)} l^{2r} \alpha_{l}^{-2r-6} 
\nonumber \\
&=& {3 \over 4} \sum_{l=0}^{\infty} \alpha_{l}^{-6}
-{15 \over 2} \sum_{l=0}^{\infty} l^{2} \alpha_{l}^{-8}
+{21 \over 2} \sum_{l=0}^{\infty}l^{4} \alpha_{l}^{-10}
\nonumber \\
&-& 2 \sum_{l=0}^{\infty} \beta_{l}(1-\beta_{l})\alpha_{l}^{-6}
+6 \sum_{l=0}^{\infty} \beta_{l} l^{2} \alpha_{l}^{-8}.
\label{(4.17)}
\end{eqnarray}

\subsection{Effect of Eq. (4.9)}

In Eq. (4.9), by virtue of the remarkable formula \cite{Moss}
\begin{equation}
\sum_{l=0}^{\infty}l^{2k}\alpha_{l}^{-2k-m}
={\Gamma \left(k+{1 \over 2}\right) \Gamma \left({m \over 2}-{1 \over 2}\right) \over
2 \Gamma \left(k+{m \over 2} \right)}x^{1-m}, \; k=1,2,3,...,
\label{(4.18)}
\end{equation}
$\Gamma$ being the standard notation for the $\Gamma$-function, we find
\begin{equation}
\sum_{l=0}^{\infty}l^{2}\alpha_{l}^{-7}={2 \over 15}h^{-4},
\label{(4.19)}
\end{equation}
a result which agrees with the application of Eq. (3.11).
Moreover, the asymptotic expansion (3.11) implies that the first sum on the 
right-hand side of (4.9) is equal to
\begin{equation}
\sum_{l=0}^{\infty}\alpha_{l}^{-5} \sim \int_{0}^{\infty}
(y^{2}+h^{2})^{-{5 \over 2}}dy+{1 \over 2}h^{-5},
\label{(4.20)}
\end{equation}
where, on defining $Y \equiv {y \over h}$, we find
\begin{equation}
\int_{0}^{\infty}(y^{2}+h^{2})^{-{5 \over 2}}dy
=h^{-4}\int_{0}^{\infty}(Y^{2}+1)^{-{5 \over 2}}dY
={2 \over 3}h^{-4}.
\label{(4.21)}
\end{equation}
It is clear once more that $h$ plays the role of regularizing parameter, since without
it the integral (4.21), and many of the integrals below, would not exist.
Last, the third sum on the right-hand side of (4.9) is again studied with the help
of (3.11), and we find
\begin{equation}
\sum_{l=0}^{\infty}{\alpha_{l}^{-5}\over (l^{2}+1)} \sim W(h)
+{1 \over 2}h^{-5},
\label{(4.22)}
\end{equation}
where
\begin{eqnarray}
W(h) & \equiv & \int_{0}^{\infty}{(y^{2}+h^{2})^{-{5 \over 2}} \over (y^{2}+1)}dy
=h^{-4}\int_{0}^{\infty}{(Y^{2}+1)^{-{5 \over 2}} \over (h^{2}Y^{2}+1)}dY 
\nonumber \\
& = & h^{-4}{\left[(2-5h^{2})\sqrt{h^{2}-1}+3h^{4}{\rm arccos}{1 \over h}\right]
\over 3 (h^{2}-1)^{5 \over 2}},
\label{(4.23)}
\end{eqnarray}
and hence no contribution to $h^{-4}$ arises from (4.22) at large $h$.

\subsection{Contribution of (4.17) and $\zeta(0)$ value}

In Eq. (4.17), by virtue of (4.18), the sums
$$
\sum_{l=0}^{\infty}l^{2} \alpha_{l}^{-8}, \;
\sum_{l=0}^{\infty}l^{4} \alpha_{l}^{-10}, 
$$
do not contribute to $h^{-4}$, while Eq. (3.11) tells us that
\begin{equation}
\sum_{l=0}^{\infty}\alpha_{l}^{-6} \sim
\int_{0}^{\infty}(y^{2}+h^{2})^{-3}dy+{1 \over 2}h^{-6}
=h^{-5}\int_{0}^{\infty}(Y^{2}+1)^{-3}dY+{1 \over 2}h^{-6},
\label{(4.24)}
\end{equation}
and hence (4.24) does not contribute to $h^{-4}$ either. Furthermore, the last two sums
on the right-hand side of (4.17), which contain the effect of $\beta_{l}$, with the
particular choice (4.6) for this coefficient are found to involve
\begin{eqnarray}
\Sigma_{\beta}^{1} & \equiv & \sum_{l=0}^{\infty}{l^{2}\over (l^{2}+1)^{2}}
\alpha_{l}^{-6} \sim \int_{0}^{\infty}{y^{2}\over (y^{2}+1)^{2}}
(y^{2}+h^{2})^{-3}dy 
\nonumber \\
& = & h^{-3} \int_{0}^{\infty}
{Y^{2}(Y^{2}+1)^{-3}\over (h^{2}Y^{2}+1)^{2}}dY
=h^{-3}{(1+4h)\pi \over 16(h+1)^{4}},
\label{(4.25)}
\end{eqnarray}
\begin{eqnarray}
\Sigma_{\beta}^{2} & \equiv & \sum_{l=0}^{\infty}{l^{2}\over (l^{2}+1)}
\alpha_{l}^{-8} \sim \int_{0}^{\infty}{y^{2}\over (y^{2}+1)}
(y^{2}+h^{2})^{-4}dy 
\nonumber \\
& = & h^{-5} \int_{0}^{\infty}
{Y^{2}(Y^{2}+1)^{-4}\over (h^{2}Y^{2}+1)}dY
=h^{-5}{[1+h(4+5h)]\pi \over 32 (h+1)^{4}},
\label{(4.26)}
\end{eqnarray}
none of which contains terms proportional to $h^{-4}$ at large h.

The contributions to $h^{-4}$ arising from all terms in (4.11) are found to vanish
with the same procedure just adopted in studying all terms in Eq. (4.17), and hence we 
find from the second line of (2.11) a contribution to $h^{-4}$ equal to
\begin{equation}
\delta_{4}=\left({9 \over 16}{2 \over 3}-{35 \over 16}{2 \over 15}\right)h^{-4}
={1 \over 12}h^{-4}.
\label{(4.27)}
\end{equation}
Eventually, we obtain from Eqs. (2.10), (3.15)-(3.17) and (4.27)
\begin{equation}
\zeta(0)=2 \left({1 \over 12}+{1 \over 4}+{1 \over 12} \right)-{1 \over 2}
={5 \over 6}-{1 \over 2}={1 \over 3},
\label{(4.28)}
\end{equation}
where $-{1 \over 2}$ is the term denoted in (2.10) by minus the contribution 
of ($l=0$), and arises from $\sigma_{2}$ in (3.6).

\subsection{Other choices of $\beta_{l}$}

For a generic choice of $\beta_{l}$ coefficient, our Eq. (4.9)  
gets replaced by
\begin{equation}
({\hat \sigma}_{4})_{I}={9 \over 16}\sum_{l=0}^{\infty}\alpha_{l}^{-5}
-{35 \over 16}\sum_{l=0}^{\infty}l^{2}\alpha_{l}^{-7}
-{3 \over 2}\sum_{l=0}^{\infty}\beta_{l}\alpha_{l}^{-5}.
\label{(4.29)}
\end{equation}
If $\beta_{l}$ is taken in the form (4.5), we find, by virtue of (3.11), the
asymptotic expansion ($K_{n}$ being the standard notation for modified Bessel 
functions of second kind and order $n$)
\begin{eqnarray}
\sum_{l=0}^{\infty}\beta_{l} \alpha_{l}^{-5} & \sim &
{1 \over \sqrt{2}}  \int_{0}^{\infty}
{e^{-{y^{2}\over 4}}\over (y^{2}+h^{2})^{5 \over 2}}dy
+{1 \over 2 \sqrt{2}}h^{-5} 
\nonumber \\
& = & {1 \over \sqrt{2}}{1 \over 48h^{2}}
e^{h^{2}\over 8}\left[h^{2}K_{0}\left({h^{2}\over 8}\right)
-(h^{2}-4)K_{1}\left({h^{2}\over 8}\right)\right]
+{1 \over 2 \sqrt{2}}h^{-5},
\label{(4.30)}
\end{eqnarray}
which has no term proportional to $h^{-4}$,
while $\beta_{l}$ in the form (4.8) leads to
\begin{equation}
\sum_{l=0}^{\infty}\beta_{l}\alpha_{l}^{-5} \sim h^{-4}
\int_{0}^{\infty}{e^{-hY}\over (Y^{2}+1)^{5 \over 2}}dY
+{1 \over 2}(1+{\tilde B}_{2})h^{-5},
\label{(4.31)}
\end{equation}
and the integral on the right-hand side of (4.31) does not have 
a term proportional to $h^{-4}$ at large $h$, either.
Moreover, the last two sums in (4.17) suggest introducing the notation
\begin{equation}
A(\beta_{l}) \equiv \sum_{l=0}^{\infty}\beta_{l}(\beta_{l}-1)\alpha_{l}^{-6}, 
\label{(4.32)}
\end{equation}
\begin{equation}
B(\beta_{l}) \equiv \sum_{l=0}^{\infty}\beta_{l}l^{2}\alpha_{l}^{-8}. 
\label{(4.33)}
\end{equation}
Now we find, for the two choices of $\beta_{l}$ according to (4.5) or (4.8),
\begin{eqnarray}
A \left({1 \over \sqrt{2}}e^{-{l^{2}\over 4}}\right)
& = & {1 \over 2}\sum_{l=0}^{\infty} e^{-{l^{2} \over 2}} \alpha_{l}^{-6}
-{1 \over \sqrt{2}} \sum_{l=0}^{\infty}
e^{-{l^{2}\over 4}} \alpha_{l}^{-6} 
\nonumber \\
& \sim & {1 \over 2} \biggr[h^{-5} \int_{0}^{\infty}
{e^{-{h^{2}Y^{2} \over 2}}\over (Y^{2}+1)^{3}}dY+{1 \over 2}h^{-6}\biggr] 
\nonumber \\
& - & {1 \over \sqrt{2}} \biggr[h^{-5} \int_{0}^{\infty}{e^{-{h^{2}Y^{2}\over 4}}
\over (Y^{2}+1)^{3}}dY+{1 \over 2}h^{-6} \biggr],
\label{(4.34)}
\end{eqnarray}
\begin{eqnarray}
A(e^{-l}) & = & \sum_{l=0}^{\infty}e^{-2l}\alpha_{l}^{-6}
-\sum_{l=0}^{\infty}e^{-l}\alpha_{l}^{-6} 
\nonumber \\
& \sim & h^{-5} \left[\int_{0}^{\infty}{{(e^{-2hY}-e^{-hY})}\over (Y^{2}+1)^{3}}dY \right]
+{{\tilde B}_{2}\over 2}h^{-6},
\label{(4.35)}
\end{eqnarray}
\begin{equation}
B \left({1 \over \sqrt{2}}e^{-{l^{2}\over 4}}\right)={1 \over \sqrt{2}}
\sum_{l=0}^{\infty}e^{-{l^{2} \over 4}} l^{2} \alpha_{l}^{-8} 
\sim {h^{-5}\over \sqrt{2}} \int_{0}^{\infty}
e^{-{h^{2}Y^{2}\over 4}}{Y^{2}\over (Y^{2}+1)^{4}}dY,
\label{(4.36)}
\end{equation}
\begin{equation}
B(e^{-l})= \sum_{l=0}^{\infty}e^{-l}l^{2}\alpha_{l}^{-8} 
\sim h^{-5} \int_{0}^{\infty}e^{-hY} {Y^{2}\over (Y^{2}+1)^{4}}dY.
\label{(4.37)}
\end{equation}
Among the integrals occurring in (4.34)-(4.37), only those on the right-hand side of (4.34) might give
rise to a contribution proportional to $h^{-4}$, because
\begin{eqnarray}
\; & \; & \int_{0}^{\infty}{e^{-{y^{2}\over 4}} \over (y^{2}+h^{2})^{3}}dy
={1 \over 64 h^{4}}\biggr[2(6-h^{2})\sqrt{\pi}
+{1\over h}(12-4h^{2}+h^{4})e^{h^{2}\over 4}\pi 
\nonumber \\
& - & (12-4h^{2}+h^{4})e^{h^{2}\over 4}\pi 
{\rm Erf} \left({h \over 2}\right) \biggr].
\label{(4.38)}
\end{eqnarray}
However, the constant coefficients within square brackets in (4.38) add up to zero,
so that no term proportional to $h^{-4}$ actually occurs at large $h$.
Such a kind of integral may be studied with the help of complex integration, because
\begin{equation}
\int_{0}^{\infty}{e^{-{y^{2}\over 4}}\over (y^{2}+h^{2})^{3}}dy
={1 \over 64}\int_{-\infty}^{\infty}
{e^{-Y^{2}}\over \left(Y^{2}+{h^{2}\over 4}\right)^{3}}dY.
\label{(4.39)}
\end{equation}
This suggests considering the following integral:
$$
\int_{\gamma}\varphi(z)dz=
\int_{\gamma}{e^{-z^{2}}\over \left(z^{2}+{h^{2}\over 4}\right)^{3}}dz,
$$
where $z=\rho e^{i \theta}$, and $\gamma$ is a rectangle with a side $\gamma_{1}$
given by the closed interval $[-\rho,\rho]$ on the real line, while the other
three sides have equation ($\varepsilon$ being a positive parameter approaching $0$,
and we take $h$ positive as well)
$$
\gamma_{2}: \; z=\rho + i \eta, \; \; \eta \in \left[0,{h \over 2}+\varepsilon \right],
$$
$$
\gamma_{3}: \; z=x+i \left({h \over 2}+\varepsilon \right), \; \;
x \in [\rho,-\rho],
$$
$$
\gamma_{4}: \; z=-\rho+i \eta, \; \;
\eta \in \left[{h \over 2}+\varepsilon,0 \right].
$$
The resulting integrand $\varphi(z)$ has a third-order pole at $z=i{h \over 2}$, with residue
\begin{equation}
\left . {\rm Res}\varphi(z) \right |_{z=i{h \over 2}}
=\left . {d^{2}\over dz^{2}}
\left[{\left(z-i{h \over 2}\right)^{3}e^{-{z^{2}}}\over 
\left(z-i{h \over 2}\right)^{3} \left(z+i{h \over 2}\right)^{3}}\right] 
\right |_{z=i{h \over 2}}
=-{i \over h^{5}}\Bigr(h^{4}-4 h^{2}+12 \Bigr) e^{h^{2}\over 4},
\label{(4.40)}
\end{equation}
which is one of the three terms occurring in (4.38).

\section{Concluding remarks}

As far as we know, our paper has performed the first $\zeta(0)$ calculation with nonlocal boundary
conditions in quantum mechanics. We have proved explicitly that at least three choices of kernel
in the nonlocal boundary operator exist (see (4.5), (4.6) and (4.8)) 
for which the $\zeta(0)$ value coincides with the
value resulting from local boundary conditions of the Robin type. Our $\zeta(0)$ value 
does not describe a one-loop conformal anomaly, as it would be the case in quantum field theory,
but it remains relevant for the understanding of scaling properties of the quantum Hamiltonian
operator.

It remains to be seen whether, for yet other choices of $\beta_{l}$ in  (4.17) and (4.29),
a contribution to $\zeta(0)$ can be found which is compatible with (1.7), (1.9), the
assumption $q \in L_{1}({\bf R}) \cap L_{2}({\bf R})$ and condition (A7) of the Appendix.
This means having to study the sums
\begin{equation}
\sum_{l=0}^{\infty}\beta_{l}\alpha_{l}^{-5}, \;
\sum_{l=0}^{\infty}\beta_{l}(\beta_{l}-1)\alpha_{l}^{-6}, \;
\sum_{l=0}^{\infty}\beta_{l}l^{2}\alpha_{l}^{-8}.
\label{(5.1)}
\end{equation}
The mere recourse to the formula \cite{Moss}
\begin{equation}
\sum_{l=0}^{\infty}l \alpha_{l}^{-1-n} \sim {h^{1-n}\over \sqrt{\pi}}
\sum_{r=0}^{\infty}{2^{r}\over r!}{\tilde B}_{r} x^{-r}
{\Gamma \left({(r+1)\over 2}\right) 
\Gamma \left({(n-1+r)\over 2}\right) \over 
2 \Gamma \left({(n+1)\over 2}\right)} 
\cos \left({r \pi \over 2}\right),
\label{(5.2)}
\end{equation}  
suggests a negative answer, because for example, upon requiring
$$
\beta_{l}(\beta_{l}-1)=\kappa \; l, \; \; \kappa >0,
$$
one finds the positive root
$$
\beta_{l}={{1 + \sqrt{1+4 \kappa l}}\over 2},
$$
which has a growth rate incompatible with (1.7) and (1.9), if one looks for functions
$q \in L_{1}({\bf R}) \cap L_{2}({\bf R})$. However, the general starting point should
be the asymptotic expansion inspired by (3.11), i.e.
\begin{equation}
\sum_{l=0}^{\infty}f(l;h) \sim \int_{0}^{\infty}f(y;h)dy+{1 \over 2}f(0;h)
+\sum_{s}{{\tilde B}_{2s}\over (2s)!}
\Bigr[f^{2s-1}(\infty;h)-f^{2s-1}(0;h)\Bigr],
\label{(5.3)}
\end{equation}
and the application of (5.3) to (5.1) deserves further work.
 
Furthermore, with the help of the experience gained from our 
calculation, it should be possible in the near future
to investigate the one-loop properties of Euclidean quantum gravity with nonlocal boundary
conditions, along the lines of Refs. \cite{Esposito99a,Esposito99b}, where it was suggested that the Universe
might become classical, after a quantum origin, by virtue of a wave function that decays as it
occurs in the case of quantum mechanical surface states \cite{Schroder} with nonlocal boundary conditions.
In order to help the general reader and stress the relevance of our work, we find it also appropriate
to write what follows.

The use of spectral $\zeta$-functions has led to several important developments over the last
decades, with application to partition functions of strings and membranes, Casimir effect, relation
between the generalized Pauli-Villars and covariant regularizations, spontaneous compactification
in two-dimensional quantum gravity, stability of the rigid membrane, topological symmetry
breaking in self-interacting theories, functional determinants for bosonic and fermionic fields,
ground-state energy under the influence of external fields, Bose-Einstein condensation of ideal
Bose gases under external conditions \cite{book1,book2,book3}. Furthermore, the work in Ref.
\cite{JMAPA} obtained heat-kernel coefficients of the Laplace operator on the $D$-dimensional 
ball, Ref. \cite{PHRVA} evaluated Casimir energies for massive fields in the bag, while the
Casimir energy for a massive fermionic quantum field with a spherical boundary was obtained
in Ref. \cite{JPAGB}. The regularization used in our paper, which relies as we said on the
pioneering work in Ref. \cite{Moss}, was applied successfully in Ref. \cite{localD} to the first
correct calculation of one-loop conformal anomaly for a massless fermionic field with local
boundary conditions, at a time when the powerful geometric formulas in Ref. \cite{Gilkey}
were not in final form. It is therefore encouraging that, after almost thirty years, such a 
regularization technique is still useful in opening yet new perspectives. 
For example, it would be interesting to apply it to the Casimir effect \cite{book3}
in cosmological backgrounds, and to establish a correspondence with yet other 
models in Ref. \cite{book3}. For the gravitational
field, one cannot generalize the Schr\"{o}der scheme by simply studying the eigenvalue problem
for an operator of Laplace type on metric perturbations $h_{\mu \nu}$, subject to nonlocal 
boundary conditions. The reason is that boundary conditions invariant under infinitesimal 
diffeomorphisms on $h_{\mu \nu}$ take, in field-theoretic language, the form \cite{CMPHA}
\begin{equation}
\Bigr[(\Pi \; h)_{ij} \Bigr]_{\partial M}=0,
\label{(5.4)}
\end{equation}
\begin{equation}
\Bigr[\Phi_{\mu}(h)\Bigr]_{\partial M}=0,
\label{(5.5)}
\end{equation}
where $\Pi$ is a projection operator that picks out only spatial components of $h_{\mu \nu}$,
while $\Phi_{\mu}(h)$ is the gauge-fixing functional. Thus, nonlocality in the boundary conditions
can only result from the gauge-fixing term; but then the invertible operator on metric perturbations
is no longer differential but it belongs to the broader class of pseudodifferential operators 
\cite{Esposito99a,Esposito99b}, for which a $\zeta(0)$ calculation is still a challenging 
task. Thus, new exciting goals are in sight in the area of physical applications of spectral
$\zeta$-functions, bearing also in mind the enlightening assessment in Ref. \cite{Witten}.

\acknowledgments
The authors are grateful to Dipartimento di Fisica ``Ettore Pancini'' for hospitality and support.

\appendix 

\section{Concepts and formulas from complex analysis}

We here summarize some concepts and results of complex analysis that are applied in our paper.

If $F$ is an entire function with a countable infinity of zeros
$\left \{ \mu_{i} \right \}$, the {\it genus} of the canonical product for $F$ 
\cite{Ahlfors} is the minimum integer $n$ such that
$$
\sum_{i=1}^{\infty}{1 \over |\mu_{i}|^{n+1}}
$$
converges. In particular, if the genus is equal to $1$, this ensures that one can write,
for some constant $\gamma$,
\begin{equation}
F(z)=\gamma z^{m}e^{g(z)} \prod_{i=1}^{\infty}
\left(1-{z \over \mu_{i}}\right)e^{z \over \mu_{i}},
\label{(A1)}
\end{equation}
where the function $g$ is entire. In particular, the function (1.6) occurring in the
nonlocal boundary condition is an entire function of genus $1$.
Such a property can be checked by pointing out that 
the zeros of $G_{l}(h)$ at large $h$ are given approximately by
\begin{equation}
h \sim l {\pi \over 2}+{\pi \over 4}+\kappa \pi, \; \kappa \in Z,
\label{(A2)}
\end{equation}
because for fixed $l$, as $h \rightarrow \infty$, one has
\begin{equation}
J_{l}(h) \sim \sqrt{2 \over \pi}{1 \over \sqrt{h}} 
\cos \left(h-l{\pi \over 2}-{\pi \over 4} \right)+{\rm O}(h^{-{3 \over 2}}),
\label{(A3)}
\end{equation}
\begin{equation}
J_{l}'(h) \sim -\sqrt{2 \over \pi}{1 \over \sqrt{h}} 
\sin \left(h-l{\pi \over 2}-{\pi \over 4} \right)+{\rm O}(h^{-{3 \over 2}}),
\label{(A4)}
\end{equation}
and hence
\begin{equation}
G_{l}(h) \sim - \sqrt{2 \over \pi} \sqrt{h} 
\sin \left(h-l{\pi \over 2}-{\pi \over 4}\right)
+{\rm O}(h^{-{1 \over 2}}),
\label{(A5)}
\end{equation}
which is independent of $\beta_{l}$. 

We further recall that, if $M(r)$ is the maximum of $F(z)$ on $|z|=r$, the {\it order} of the entire
function $F$ is defined to be \cite{Boas}
\begin{equation}
{\rm ord}(F) \equiv \lim_{r \to \infty} \; {\rm sup} \;
{\log \; \log(M(r)) \over \log(r)}.
\label{(A6)}
\end{equation}
An example of an entire function of order $n$ is $e^{z^{n}}$. Our function $G_{l}$ in (1.6)
can be also studied from the point of view of its order, as defined in (A6), and its relation
with the genus \cite{Boas}.

Moreover, we can apply to Eq. (1.6) a theorem studied, among the others, by Watson
\cite{Watson}, according to which, if $A$ and $B$ are real and $\nu > -1$, the function
$AJ_{\nu}+Bz J_{\nu}'$ has all its zeros real, except that it has $2$ purely imaginary
zeros when ${A \over B}+\nu <0$. This implies that our $G_{l}(h)$ has only real
roots if 
\begin{equation}
\beta_{l}+l \geq 0.
\label{(A7)}
\end{equation}
This condition is fulfilled by all forms (4.5)-(4.8) of $\beta_{l}$ considered
in our paper.

The uniform asymptotic expansions of Bessel functions and their first derivative 
are due to Debye and Olver \cite{Olver}, and are used extensively in Ref. \cite{Moss} and in our
paper. They read as
\begin{equation}
J_{l}(ih) \sim {(ih)^{l}\over \sqrt{2 \pi}} \alpha_{l}^{-{1 \over 2}}
\; e^{\alpha_{l}}
\; e^{-l \log(l+\alpha_{l})} \Sigma_{1},
\label{(A8)}
\end{equation}
\begin{equation}
J_{l}'(ih) \sim {(ih)^{l-1}\over \sqrt{2 \pi}} \alpha_{l}^{{1 \over 2}}
\; e^{\alpha_{l}}
\; e^{-l \log(l+\alpha_{l})} \Sigma_{2},
\label{(A9)}
\end{equation}
where (see also (2.13)-(2.19))
\begin{equation}
\Sigma_{1} \sim u_{0}+{u_{1}\over l}+{u_{2}\over l^{2}}
+{u_{3}\over l^{3}}+...,
\label{(A10)}
\end{equation}
\begin{equation}
\Sigma_{2} \sim v_{0}+{v_{1}\over l}+{v_{2}\over l^{2}}
+{v_{3}\over l^{3}}+...,
\label{(A11)}
\end{equation}
and, in particular, 
\begin{equation}
u_{3}(\tau)={1 \over 2}\tau^{2}(1-\tau^{2})u_{2}'(\tau)
+{1 \over 8}\int_{0}^{\tau}(1-5 \rho^{2})u_{2}(\rho)d\rho,
\label{(A12)}
\end{equation}
\begin{equation}
v_{3}(\tau)=u_{3}(\tau)+\tau(\tau^{2}-1) \left[{1 \over 2}u_{2}(\tau)+\tau u_{2}'(\tau) \right].
\label{(A13)}
\end{equation}

\end{document}